\documentclass[10pt]{iopart}
\usepackage{iopams}
\usepackage{amssymb}
\usepackage{graphicx}
\usepackage{color}
\usepackage[tight,TABTOPCAP]{subfigure}
\usepackage{epstopdf}
\begin{document}
\makeatletter

\newbox\slashbox \setbox\slashbox=\hbox{$/$}
\newbox\Slashbox \setbox\Slashbox=\hbox{$/$}
\def\pFMslash#1{\setbox\@tempboxa=\hbox{$#1$}
  \@tempdima=0.5\wd\slashbox \advance\@tempdima 0.5\wd\@tempboxa
  \copy\slashbox \kern-\@tempdima \box\@tempboxa}
\def\pFMSlash#1{\setbox\@tempboxa=\hbox{$#1$}
  \@tempdima=0.5\wd\Slashbox \advance\@tempdima 0.5\wd\@tempboxa
  \copy\Slashbox \kern-\@tempdima \box\@tempboxa}
\def\FMslash{\protect\pFMslash}
\def\FMSlash{\protect\pFMSlash}
\def\miss#1{\ifmmode{/\mkern-11mu #1}\else{${/\mkern-11mu #1}$}\fi}
\makeatother


\title[Revisiting the flavor changing...]{Revisiting the flavor changing neutral current Higgs decays $H\to q_i q_j$ in the Standard Model}
\author{L. G. Ben\'\i tez-Guzm\' an$^{(a)}$, I. Garc\'\i a-Jim\' enez$^{(b)}$, M. A. L\' opez-Osorio$^{(a)}$, E. Mart\'\i nez-Pascual$^{(a)}$, J. J. Toscano$^{(a)}$}
\address{$^{(a)}$Facultad de Ciencias F\'{\i}sico Matem\'aticas,
Benem\'erita Universidad Aut\'onoma de Puebla, Apartado Postal
1152, Puebla, Puebla, M\'exico.\\
$^{(b)}$Instituto de F\'{\i}sica, Benem\'erita Universidad Aut\'onoma de Puebla. Apdo. Postal J-48, C.P. 72570 Puebla, Pue.,
M\'exico.}

\begin{abstract}
An exact calculation of the Higgs boson decays $H\to q_i q_j$ mediated by flavor changing neutral currents is presented in the context of the Standard Model. Using up-to-date experimental data, branching ratios of the order of  $10^{-7}$, $10^{-8}$, $10^{-8}$, and $10^{-15}$ are found for the $\bar{b}s+\bar{s}b$, $\bar{b}d+\bar{d}b$, $\bar{s}d+\bar{d}s$, and $\bar{c}u+\bar{u}c$ decay modes, respectively.
\end{abstract}



\section{Introduction}
\label{I}
In the light of the observation of a Standard Model (SM)-like Higgs boson~\cite{ATLAS,CMS}, some rare or suppressed processes of this particle become relevant. The Higgs boson has a relatively small decay width, and this is a good reason to revisit some of its rare decays, as the corresponding branching ratios may be significantly enhanced. Naturally suppressed (one-loop) decays of this particle are the flavor changing neutral currents (FCNC) decays $H\to q_iq_j$, with $H$ the Higgs boson and $q_iq_j=bs,\, bd,\, sd,\, cu$. In these processes the corresponding amplitudes crucially depend on the quotient $m^2_k/m^2_W$, with $m_k$ the mass of the quark circulating around the loop.  For $m^2_k/m^2_W\ll 1$, the corresponding decays are strongly suppressed; this is known as the Glashow-Illiopolus-Maiani (GIM) mechanism~\cite{GIM}. Examples of such decays are the following FCNC transitions of the top quark: $t\to cV$, with $V=\gamma, \, g, \, Z$ (see~\cite{ES}), and $t\to cH$ (see~\cite{MP}). However, no GIM suppression is present for $m^2_k/m^2_W> 1$; this is the case for the well-known FCNC $b\to s\gamma$ decay, whose branching ratio is ten orders of magnitude higher of the aforementioned top quark decays. This suggests that the FCNC Higgs decays into quarks of down type, especially the $bs$ mode, could have interesting branching ratios. Some features of these decays have previously been studied in the context of the SM. The $q_iq_jH$ vertex was studied in~\cite{SM} using some approximations. Several authors~\cite{BSH} explored the possibility of a very light Higgs boson via the $b\to sH$ decay. Also, some technical aspects of the one-loop $sdH$ vertex were studied in~\cite{BL}. This problem has also been studied in some SM extensions: the SM with a fourth generation~\cite{SM4G}, the Two Higgs Doublet Model~\cite{THDM} and the Minimal Supersymmetric Standard Model~\cite{MSSM}. The purpose of this paper is to present exact formulae for the branching ratios of the FCNC $H\to q_iq_j$ decays within the context of the SM and numerically analyze them with the use of up-to-date experimental data.

The paper has been organized as follows. In section \ref{C}, exact formulae for the $H\to q_iq_j$ decays are derived in the context of the SM. Section \ref{D} is devoted to discuss our results. Finally, in section \ref{Con} the conclusions are presented.

\section{The $H\to q_iq_j$ decays}
\label{C}

\begin{figure}[h]
\centering\includegraphics[scale=.7]{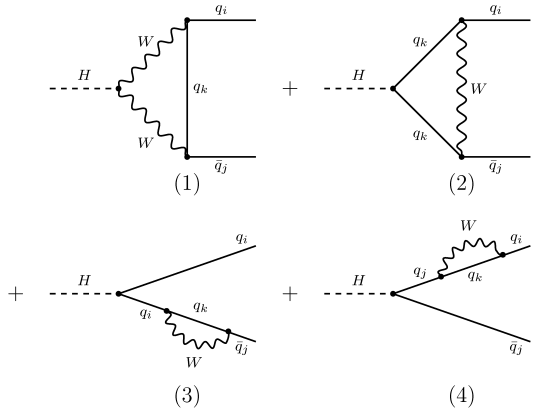}
\caption{\label{FIG1}Diagrams contributing to the $H\to q_iq_j$ decay in the unitary gauge.}
\end{figure}

In the unitary gauge, the $H\to q_j q_i$ decay arises through the one-loop diagrams shown in figure \ref{FIG1}. The corresponding invariant amplitude is given by
\begin{equation}
{\cal M}\left(H\to  q_j q_i\right)=-\frac{i}{(4\pi)^2}\,\frac{g^3}{2}\, \bar{u}(p_i,s_i)\left(F_L\,P_L+F_R\, P_R\right)v(p_j,s_j)\, ,
\end{equation}
where the projection operators are $P_{L,R}=(1\mp \gamma_5)/2$, and the right loop amplitude $ F_{R} $ is related to the left loop amplitude $F_{L}$ via:
\begin{equation}\label{leftright}
F_R=F_L(m_i \longleftrightarrow m_j)\, .
\end{equation}
Here
\begin{equation}
\label{FL}
F_L= \sqrt{\frac{x_i}{x_W}}\,\sum_k\, V_{ik}V^*_{kj}\left[f_0+\sum^2_{l=1}f_l\, m^2_H\, C_0(l)+\sum^7_{l=1}g_l\, B_0(l)\right]\, ,
\end{equation}
where $V$ is the Cabibbo-Kobayashi-Maskawa (CKM) matrix, whereas $C_0(l)$ and $B_0(l)$ are Passarino-Veltman scalar functions given by
\begin{eqnarray}
C_0(1)&=&C_0(m^2_i,m^2_j,m^2_H,m^2_W,m^2_k,m^2_W)\, ,\\
C_0(2)&=&C_0(m^2_i,m^2_j,m^2_H,m^2_k,m^2_W,m^2_k)\, ,
\end{eqnarray}
\begin{eqnarray}
\label{b1}
B_0(l)&=&B_0(m^2_l,m^2_k,m^2_W)\, , \, \, l=1,2 ; \, \, m_1=m_i,\, m_2=m_j\, ,\\
\label{b2}
B_0(r)&=&B_0(m^2_H,m^2_r,m^2_r)\, , \, \, r=3,4 ; \, \, m_3=m_W, \, m_4=m_k\, ,\\
\label{b3}
B_0(5)&=&B_0(0,m^2_k,m^2_W)\, ,\\
\label{b4}
B_0(s)&=&B_0(0,m^2_s,m^2_s)\, , \, \, s=6,7; \, \, m_6=m_W, m_7=m_k \, .
\end{eqnarray}
The masses $m_i$ and $m_j$ are the external quarks' masses, whereas $m_k$ denotes the mass of a quark circulating in the loop.

The correct implementation of the GIM mechanism requires that, before evaluating the loop amplitudes, those terms that do not depend on the internal quark mass must be removed, that is, terms of the form
\begin{equation}
\sum_k\, V_{ik}V^*_{kj}\,f(m_i,m_j,m_H,m_W)\, ,
\end{equation}
with $f(m_i,m_j,m_H,m_W)$ being arbitrary functions which do not depend on the internal mass $m_k$ and which vanish for $i\neq j$ due to the unitarity of the $V$ matrix. Taking into account this fact, we can write the various form factors appearing in the loop amplitudes as follows:
\begin{equation}
{f_0} = \frac{{{x_k}\left( {{x_k} + {x_W} - {x_j}} \right)}}{{2{x_W}\left( {{x_j} - {x_i}} \right)}}\, ,
\end{equation}
\begin{eqnarray}
{f_1} &=& \frac{1}{2x_W\delta}\bigg  \{     {x_W}\Big[ 2\left( {{x_j} - {x_i}} \right)\left( {x_i^2 - {x_i}{x_j} + x_j^2} \right) \nonumber \\
&&+2{x_i}\left( {2{x_i} - 1} \right) - {x_j}\left( {{x_i} + {x_j} + 1} \right)\Big]\nonumber \\
&&+ \left( {{x_i} - {x_j} - 1} \right)     \Big [    2x_W^2\left( {2{x_W} - 1} \right) \nonumber \\
&&+ {x_W}{x_k}\left( {1 + 2\left( {{x_i} + {x_j} - {x_W} - {x_k}} \right)} \right)\nonumber \\
&&+ {x_k}\left( {{x_i} + {x_j} - {x_k} - 1} \right) - {x_i}{x_j}   \Big]\nonumber \\
&&+ 2{x_W^2}{x_j}\left( {{x_i} - {x_j} + 1} \right)  \bigg  \}\, ,
\end{eqnarray}
\begin{eqnarray}
{f_2} &=& \frac{x_k}{2x_W\delta} \Big \{       2{x_W}\left[ {{x_i} - \left( {{x_i} - {x_j}} \right)\left( {{x_i} + 2{x_j}} \right)} \right]\nonumber \\
&& + {x_k}\left[ {{{\left( {{x_i} - {x_j}} \right)}^2} - {x_i} - 3{x_j}} \right]     \nonumber \\
&\,&+ \left( {{x_i} - {x_j} - 1} \right)\left[ {4x_W^2 - 2x_k^2 + {x_W}\left( {1 - 2{x_k}} \right)} \right] \nonumber \\
&&+ {x_j}\left( {{x_i} + {x_j}} \right)\left( {{x_i} - {x_j} + 1} \right)     \Big  \}\, ,
\end{eqnarray}
\begin{eqnarray}\fl
  &{g_1} =  \frac{1}{4{x_W}{x_i}\left( {{x_i} - {x_j}} \right)\delta} \bigg \{     \left( {2x_W^2 - {x_W}{x_k} - x_k^2} \right)\nonumber \\
&\times \Big [   4x_i^3 + x_i^2\left( {{x_j} - 4} \right) - 2{x_i}{x_j}\left( {3{x_j} - 1} \right)+ {x_j}{{\left( {{x_j} - 1} \right)}^2}    \Big ] \nonumber \\
&- {x_W}{x_i}\left[ {4\left( {x_i^3 - x_j^3} \right) - x_i^2\left( {3{x_j} + 4} \right) + 2{x_i}{x_j}\left( {{x_j} - 1} \right) + {x_j}{{\left( {{x_j} + 1} \right)}^2}} \right]   \nonumber\\
&\,+ 2{x_k}{x_i}\left[ {{x_i}\left( {{{\left( {{x_i} + {x_j}} \right)}^2} - 4x_j^2 - 2{x_j} - 1} \right) - 2{x_j}\left( {{x_j} - 1} \right)} \right]\nonumber \\
&  - {x_j}x_i^2\left[ {{{\left( {{x_i} - {x_j}} \right)}^2} + 2{x_i} - 6{x_j} + 1} \right]      \bigg \}\, ,
\end{eqnarray}
\begin{eqnarray}
{g_2} &=& \frac{1}{4{x_W}\left( {{x_i} - {x_j}} \right)\delta}  \bigg  \{       \left( {x_k^2 + {x_W}{x_k} - 2x_W^2} \right)\nonumber \\
&&\times \left[ {\left( {{x_i} - {x_j}} \right)\left( {{x_i} + 7{x_j}} \right) - 2\left( {{x_i} + {x_j}} \right) + 1} \right]      \nonumber \\
&\,& + {x_W}\left[ {3{x_i}{{\left( {{x_i} - 1} \right)}^2} + {x_i}{x_j}\left( {7{x_j} - 4{x_i} + 2} \right) - 2{x_j}\left( {3x_j^2 + 1} \right)} \right] \nonumber \\
&& - {x_k}\big[ {x_i}{{\left( {{x_i} - 1} \right)}^2} + {x_i}{x_j}\left( {3{x_i} - {x_j}} \right) + {x_j}\left( {1 - 3x_j^2 - 6{x_j}} \right) \big]\nonumber \\
 &&- {x_j}\left[ {{x_i}\left( {1 - x_i^2} \right) + {x_i}{x_j}\left( {2{x_i} - {x_j} + 2} \right) + 2{x_j}\left( {{x_j} - 1} \right)} \right]      \bigg \}\, ,
\end{eqnarray}
\begin{eqnarray}
{g_3} &=& \frac{x_k}{2x_W\delta}\left( 2x_W+ 1 \right)\left( 1-x_i + x_j \right)\, , \\
{g_4}&=& \frac{x_k}{2x_W\delta}\Big[ 2\left( x_k + 2x_W \right)\left( x_i - x_j - 1 \right)\nonumber \\
&&- 2x_j\left( x_i - x_j \right) - x_i - x_j + 1 \Big]\, , \\
{g_5} &=& \frac{{\left( {{x_i} + {x_j}} \right)\left( {{x_W} - {x_k}} \right)\left( {{x_i} - {x_k} - 2{x_W}} \right)}}{{4{x_i}{x_W}\left( {{x_i} - {x_j}} \right)}}\, ,\\
{g_6} &=& \frac{{{x_k}}}{{2\left( {{x_i} - {x_j}} \right)}}\, ,\\
{g_7} &=& \frac{{{x_k}\left( {{x_j} - {x_k} - 2{x_W}} \right)}}{{2{x_W}\left( {{x_i} - {x_j}} \right)}}\, ,
\end{eqnarray}
where we have introduced the dimensionless variables $x_i\equiv m^2_i/m^2_H$,  $x_j\equiv m^2_j/m^2_H$,  $x_k\equiv m^2_k/m^2_H$, and  $x_W\equiv m^2_W/m^2_H$. In addition, $\delta=1-2(x_i+x_j)+(x_i-x_j)^2$. In $ f_{0} $, terms which do not depend on $m_k$ have been removed. We proceeded in the same fashion for the form factors $g_3$ and $g_6$, as the associated $B_0(3)$ and $B_0(6)$ functions do not depend on $m_k$. It turns out that the loop amplitudes are free of ultraviolet divergences, as it can be verified by adding together the form factors associated with the $B_0(l)$ functions. This leads to
\begin{equation}
\sum^7_{l=1}g_l=\frac{x_i-x_j}{4x^2_W}\, ,
\end{equation}
which vanishes after using the unitarity of the CKM matrix. In this respect, the numerical evaluation of the $B_0$ functions must be performed with some care, as these functions may still contain a $m_k$-independent part, which could lead to additional cancellations due to the GIM mechanism. This means that we cannot use software such as FF~\cite{FF} or LoopTools~\cite{LT}, to directly perform such evaluation. Instead, we use their analytic solutions in order to remove any redundant contribution to the loop amplitudes. As it can be seen from Eqs.~(\ref{b1}) to (\ref{b4}), they include three distinct $B_0$ functions: (i) In Eq.~(\ref{b1}), $ B_{0}(1) $ and $ B_{0}(2) $ have an analytic solution, namely
\begin{equation}\fl
B_0(l)=\Delta+1+\frac{x_W}{x_k-x_W}\log(x_W)-\frac{x_k}{x_k-x_W}\log(x_k)+F(x_l,x_k,x_W)\, ,
\end{equation}
where $l=1,2$, $x_{(1,2)}=x_{(i,j)}$, and $\Delta=\frac{1}{\epsilon}-\gamma_E-\log\left(\frac{m^2_H}{4 \pi \mu^2}\right)$ is a divergent factor, which is common to all the $B_0$ functions and thus vanishes in the total amplitude, since this is free of ultraviolet divergences. In addition, the function $F(x_l,x_k,x_W)$ is
\begin{eqnarray}\fl 
\qquad F(x_l,x_k,x_W)=&1+\frac{1}{2}\left(\frac{x_k-x_W}{x_l}-\frac{x_k+x_w}{x_k-x_W} \right)\log\left(\frac{x_W}{x_k}\right)\nonumber \\
&+\frac{x_+x_-}{x_l}\log\left(\frac{x_++x_-}{x_+-x_-}\right)\, , \, \, \, \, x_l<(\sqrt{x_k}-\sqrt{x_W})^2 \,,
\end{eqnarray}
where $x_\pm\equiv \sqrt{\left(\sqrt{x_k}\pm \sqrt{x_W}\right)^2-x_l}$. Notice that $F(x_l,x_k,x_W)$ vanishes in the limit $x_l \to 0$. (ii) In Eq.~(\ref{b2}),  functions $B_0(3)$ and $ B_{0}(4) $ also have an analytic solution which is
\begin{equation}
B_0(r)=\Delta+2-\log(x)-F(x)\, ,
\end{equation}
where $r=3,4$, the variable $x$ stands for $x_k$ or $x_W$, and
\begin{equation}
F(x)=2\sqrt{4x-1}\arctan\left(\frac{1}{\sqrt{4x-1}}\right)\, , \ \ \  x>1/4\, .
\end{equation}
Finally, (iii)  functions $B_0(5)$, $ B_{0}(6) $ and $ B_{0}(7) $ in Eqs.(\ref{b3}) and (\ref{b4}) become
\begin{eqnarray}
B_0(5)&=& \Delta +1 +\frac{x_W}{x_k-x_W}\log(x_W)-\frac{x_k}{x_k-x_W}\log(x_k)\, , \\
B_0(6,7)&=&\Delta-\log(x_W,x_k)\, .
\end{eqnarray}

Once plugging the expressions for the $B_0$ functions into the $F_L$ amplitude,
\begin{eqnarray}
F_L&=& \sqrt{\frac{x_i}{x_W}}\,\sum_k\, V_{ik}V^*_{kj}\Big[\bar{f}_0+\sum^2_{l=1}f_l\, m^2_H\, C_0(l)\nonumber \\
&&+f_3\, \log\left(\frac{x_W}{x_k}\right)
+f_4 \, F(x_W)+f_5\, F(x_k)\nonumber \\
&&+f_6\, F(x_i,x_k,x_W)+f_7\, F(x_j,x_k,x_W)\Big]\, ,
\end{eqnarray}
where
\begin{equation}
\bar{f}_0=\frac{x_k}{x_W\,\delta}\left[(x_k+x_W)(1-x_i+x_j)+x_i-1+x_j(x_i-x_j)\right]\, ,
\end{equation}
\begin{equation}\fl
\qquad \quad \ \  f_3=\frac{x_k}{2x_W(x_k-x_W)\delta}\, \left[(x_k-x_W)(x_i-x_j-1)+x_j(x_j-x_i-1)\right]\, ,
\end{equation}
\begin{equation}
f_4=\frac{x_k}{2x_W\delta}(2x_W+1)(x_i-x_j-1)\, ,
\end{equation}
\begin{equation}\fl
\qquad \quad \ \ \ f_5=\frac{x_k}{2x_W\delta}\left[2(x_k+2x_W)(1-x_i+x_j)+2x_j(x_i-x_j)+x_i+x_j\right]\, ,
\end{equation}
\begin{eqnarray}\fl
\qquad \quad \ \ \ f_6=&&\frac{x_k}{4x_ix_WD}\Big\{(x_k+x_W)\nonumber \\
&&\times \left[x_i\left( 4x_i(1-x_i)-x_ix_j+2x_j(3x_j-1)\right)-x_j(1-x_j)^2 \right]\nonumber \\
&&+2x_i\left[x_i\left((x_i+x_j)^2-2x_j(2x_j+1)-1\right)+4x_j(1-x_j)\right] \Big\}\, ,
\end{eqnarray}
\begin{eqnarray}
f_7&=&\frac{x_k}{4x_WD}\Big\{(x_k+x_W)\left[(1-x_i-x_j)^2+4x_j(x_i-2x_j)\right]\nonumber \\
&&+x_i\left[x^2_j-(1-x_i)^2\right]-3x_j\left[x^2_i-(1+x_j)^2+\frac{4}{3}\right] \Big\}\, ,
\end{eqnarray}
where $D=(x_i-x_j)^3-2(x^2_i-x^2_j)+x_i-x_j$. The $F_L$ amplitude is now free of redundances. The right loop amplitude is obtained from  $F_R=F_L(x_i\leftrightarrow x_j)$ [\emph{cf.}~Eq.~(\ref{leftright})].

\section{Discussion}
\label{D}

It can be verified that our results satisfy the following consistency conditions: $(i)$ The amplitudes for the diagrams (1) and (2) vanish for $m_i=0=m_j$ (diagrams (3) and (4) do not exist in this case); $(ii)$ loop amplitudes $F_L$ and $F_R$ are free of ultraviolet divergences once the unitarity of the CKM matrix is used; $(iii)$ in the limit  $m_j\to 0$ ($m_i\to 0$) the right (left) loop amplitude $F_R$ ($F_{L}$) vanishes and the left (right) loop amplitude is different from zero.

We  now explore the behavior of the loop amplitudes in the heavy mass limit. Let $m_W=m_k$ in the expression for $F_L$ given by Eq.(\ref{FL}), in this limit we have $C_0(1)=C_0(2)=C_0(m^2_i,m^2_j,m^2_H,m^2_k,m^2_k,m^2_k)$, $B_0(3)=B_0(4)=B_0(m^2_H,m^2_k,m^2_k)$, and $B_0(5)=B_0(6)=B_0(7)=B_0(0,m^2_k,m^2_k)$. Therefore
\begin{eqnarray}
F_L|_{m_W\to m_k}&=&\sum_{k}V_{ik}V^*_{kj}\Big\{-\frac{1}{4}\sqrt{\frac{x_i}{x_k}}\left[2+\frac{\log(x_k)}{x_k}\right]+c_{12}f(x_k)\nonumber \\
&&+b_1F(x_i,x_k)+b_2F(x_j,x_k)+b_{34}F(x_k)\Big\}\, ,
\end{eqnarray}
where
\begin{eqnarray}
c_{12}&=&\frac{1}{2}\sqrt{\frac{x_i}{x_k}}\left[\frac{x_ix_j(1-x_i+x_j)}{x_k}+x_k+1-2x_i+x_j\right]\, , \\
b_1&=&-\frac{1}{4(x_i-x_j)}\sqrt{\frac{x_i}{x_k}}\Big[\frac{x_ix_j(\delta +4(x_i-x_j))}{\delta\, x_k}\nonumber \\
&&+2(x_i-x_j)-x_j\Big]\, ,\\
b_2&=&\frac{1}{4(x_i-x_j)}\sqrt{\frac{x_i}{x_k}}\Big[\frac{x_ix_j(\delta+2x_i)+2x^2_j(1-x_j)}{\delta\, x_k}\nonumber \\
&&+2(x_i-x_j)-x_j \Big]\, ,\\
b_{34}&=&\frac{1}{2}\sqrt{\frac{x_i}{x_k}}\left[\frac{(x_i-1)^2-x^2_j}{\delta \, x_k}+2\right]\, .
\end{eqnarray}
In addition,
\begin{eqnarray}
\label{C0A}
f(x_k)&=&-2\left[\arcsin\left(\frac{1}{\sqrt{x_k}}\right)\right]^2\, ,\\
F(x,x_k)&=&i\sqrt{\frac{4x_k}{x}-1}\, \log\left(\frac{\sqrt{\frac{4x_k}{x}-1}+i}{\sqrt{\frac{4x_k}{x}-1}-i}\right)\, , \, \, \, x=x_i,x_j \, ,
\end{eqnarray}
where $f(x_k)$ is the solution of the $m^2_HC_0(m^2_i,m^2_j,m^2_H,m^2_k,m^2_k,m^2_k)$ function for $m_i=0=m_j$, provided that this function depends weakly on these masses. From these expressions, it is easy to show that each term within $F_L|_{m_W\to m_k}$ separately vanishes for $x_k \to \infty$.

Let us now provide the calculations for the branching ratio of the $H\to \bar{q}_j q_i+\bar{q}_iq_j$ decays. In general, one has
\begin{eqnarray}
BR(H\to \bar{q}_j q_i+\bar{q}_iq_j)&=&\left(\frac{N_C}{128\pi^2}\right)\left(\frac{\alpha}{s_W}\right)^3\left(\frac{m_H}{\Gamma_H}\right)\sqrt{\delta}\nonumber \\
&&\times
\Big[(1-x_i-x_J)\left(|F_L|^2+|F_R|^2\right)\nonumber \\
&&\, \, \, \, \, \, \, \, -4\sqrt{x_ix_j}Re\left(F_LF^*_R\right)\Big]\, ,
\end{eqnarray}
where $N_C=3$ is the color index and we have added a factor of 2 in order to consider both, $\bar{q}_jq_i$ and $\bar{q}_iq_j$, possibilities. To evaluate all the branching ratios of the allowed decays, the following values for the Higgs boson mass and decay width  were used: $m_H=125 \, GeV$ and $\Gamma_H=4.403\times 10^{-3}\, GeV$; the values for the remaining parameters involved are those reported by the Particle Data Group~\cite{PDG}.

 We evaluated the three point scalar functions $C_0(1)$ and $C_0(2)$ using FF~\cite{FF} and LoopTools~\cite{LT}. Also, we perform an independent numerical evaluation of these functions by starting from their integral representation. In terms of Feynman parameters, the $C_0(1)$ function can be written in the following form:
\begin{equation}\fl
\label{C0E}
\qquad \quad m^2_HC_0(1)=\int^1_0 dx \frac{2}{\sqrt{\bar{\Delta}}}\left[\arctan\left(\frac{(2-a)x-1}{\sqrt{\bar{\Delta}}}\right)-
\arctan\left(\frac{1-ax}{\sqrt{\bar{\Delta}}}\right)\right]\, ,
\end{equation}
where $a=1+x_i-x_j$ and
\begin{eqnarray}
\bar{\Delta}&=&4x_W-1+2\left(1-x_i-x_j+2(x_k-x_W)\right)x-\delta\, x^2 \, .
\end{eqnarray}
Some simplifications are obtained in the limit $x_i=0=x_j$. In this case, the above integral reads as follows:
\begin{equation}
m^2_HC_0(1)=-\int^1_0 dx \frac{4}{\sqrt{\hat{\Delta}}}\arctan\left(\frac{(1-x}{\sqrt{\hat{\Delta}}}\right)\, ,
\end{equation}
where
\begin{equation}
\hat{\Delta}=4x_W-1+2\left(1+2(x_k-x_W)\right)x-x^2\, .
\end{equation}
However, as it occurs in the exact case, this integral cannot be expressed in terms of elementary functions. If in addition, one assumes that $x_W=x_k$, the result (\ref{C0A}) is obtained. On the other hand, the corresponding expression for $m^2_HC_0(2)$ is obtained from $m^2_HC_0(1)$ via the interchange $x_k\leftrightarrow x_W$. The numerical evaluation of the integral given by Eq. (\ref{C0E}) leads to results that are in excellent agreement with those obtained using the FF and LoopTools programs. 

The branching ratios have been evaluated using the following approximations. For those decays of type down, only the contribution of the quark top was considered for the channels $bs$ and $bd$, whereas for the $ds$ channel, only the contribution of the $c$ quark was taken into account. It results that in the case of the $ds$ channel, the $t$ contribution is quite marginal with respect to the $c$ contribution due to a strong suppression factor coming from the CKM matrix. While the function $F_L$ ($F_R$) induced by the $t$ quark is one order of magnitude (of the same order of magnitude) with respect to the $c$ contributions, the CKM effects are, respectively, $|V_{st}V_{td}|=3.36\times 10^{-4}$ and $|V_{sc}V_{cd}|=0.222$. As far as the $cu$ channel is concerned, only the contribution of the $b$ quark was included.

Our results are displayed in Table.\ref{TABLE1}. The values of $m^2_H\, C_0(1)$ and $m^2_H\, C_0(2)$ are shown in Table.\ref{TABLE2}. From Table.\ref{TABLE1}, we can see that the FCNC Higgs decays into the $bs$, $bd$, $sd$, and $cu$ modes have corresponding branching ratios of $3\times 10^{-7}$, $1.14\times 10^{-8}$, $1.2\times 10^{-8}$, and $5\times 10^{-15}$. It is worth comparing these branching ratios with those associated with the decays $b\to s\gamma$, $t\to cg$, $t\to c\gamma$, $t\to cZ$, and $t\to cH$, which are of the order of $10^{-4}$, $4\times 10^{-11}$,  $5\times 10^{-13}$, $10^{-13}$, and $10^{-14}$, respectively. Although significant, compared with the FCNC top quark transitions, the $H\to bs$ decay is quite suppressed to be detected in future experiments.

\begin{table}[!ht]
\centering
\caption{\label{TABLE1}
Branching ratios for the FCNC Higgs decays $H\to q_iq_j$. No approximations were made.}
\begin{tabular}{|l|c|c|c|c|c|c|}\hline\hline
$H\to q_iq_j$  \ & \ $|V_{ik}V_{kj}|^2$ \ & \ $F_L$ \ & \ $F_R$ \ & \ $BR$ \\ \hline
$H\to bs$\ & \  $1.66\times 10^{-3}$ \ & \ $-0.2246$  \ & $0.1769$ \ & $2.92\times 10^{-7}$ \\ \hline
$H\to bd$\ & \  $7.34\times 10^{-5}$ & \ $-0.2247$  \ & $0.1469$ \ & $1.14\times 10^{-8}$  \\ \hline
$H\to sd$\ & \ $4.9\times 10^{-2}$ \ & $7.95\times 10^{-4}$  \ & $-1.06\times 10^{-2}$ \ & $1.2\times 10^{-8}$ \\ \hline
$H\to cu$ \ & \ $2.88\times 10^{-8}$ \ & $(-3.8 + 0.047i)\times 10^{-3}$ \ & $-8.41\times 10^{-3}$ \ & $5.31\times 10^{-15}$\\
\hline\hline
\end{tabular}
\end{table}

\begin{table}[!ht]
\centering
\caption{\label{TABLE2}
Values of the diverse scalar $C_0$ functions.}
\begin{tabular}{|l|c|c|c|c|c|c|}\hline\hline
$H\to q_iq_j$ \ & \ $m^2_H\,C_0(1)$ \ & \ $m^2_H\,C_0(2)$  \\ \hline
$H\to bs$ \ & \ $-0.79916758$  \ & \ $-0.40298304$ \\ \hline
$H\to bd$ \ & \ $-0.79916754$  \ & \ $-0.40298302$ \\ \hline
$H\to sd$ \ & \ $-3.17017656$  \ & \ $-0.58-3.86i$ \\ \hline
$H\to cu$ \ & \ $-3.13929925$  \ & \ $-0.61 -3.85i$ \\
\hline\hline
\end{tabular}
\end{table}

Our results should be compared with recent constraints for flavor violating Higgs decays derived from low-energy data. By assuming a renormalizable general effective Lagrangian for the Yukawa sector, bounds on lepton and quark flavor violating decays of the Higgs boson were analyzed in references~\cite{BEI,HKZ}. In particular, experimental limits on $B_s$ physics were used to derive constraints on the $H\to bs$ decay. The authors of these papers found that $BR(H\to bs)<4\times 10^{-4}$~\cite{BEI} and $BR(H\to bs)<2\times 10^{-3}$~\cite{HKZ}. Although away from the SM prediction found here, these branching ratios would even be out of the reach of the LHC due to large QCD background.

\section{Conclusions}
\label{Con}

In conclusion, we have presented exact formulae for the FCNC Higgs decays $H\to q_iq_j$ in the context of the SM. Recent experimental data were used to predict the branching ratios for all the kinematic allowed modes. Although the branching ratios for the decay $H\to bs$ are significantly larger than the ones associated with the top quark transitions $t\to u_i\gamma$, $t\to u_iZ$, and $t\to u_iH$, our numerical analysis suggests that these decays are out of reach of future experiments, and thus they may be very sensitive to new physics effects. Recent analysis using experimental data from $B_s$ physics shows that branching ratios up to four orders of magnitude larger than the SM prediction could be allowed for $BR(H\to bs)$; however, they would still be undetectable at the LHC~\cite{BEI,HKZ}. We think that our results can be useful for people interested in investigating these decays in other contexts of new physics.

\section*{We acknowledge financial support from CONACYT. M.~A.~L.-O.,  E.~M.-P. and J.~J.~T. also acknowledge SNI (M\' exico).}

\end{document}